\RequirePackage{fix-cm}
\documentclass[smallextended]{svjour3}       
\smartqed  
\usepackage{graphicx}
\usepackage{CJK}
\usepackage{indentfirst}
\usepackage{bm}
\usepackage{amsmath}
\usepackage{amsfonts}
\newcommand{\be}{\begin{equation}}
\newcommand{\ee}{\end{equation}}
\newcommand{\ba}{\begin{array}}
\newcommand{\ea}{\end{array}}
\journalname{Int J Theor Phys}
\begin{document}

\title{Unextendible maximally entangled bases and mutually unbiased bases in multipartite systems 
}

\author{Ya-Jing Zhang    \and
        Hui Zhao         \and
        Naihuan Jing    \and
        Shao-Ming Fei 
}

\institute{Y.J. Zhang \and H. Zhao \at
             College of Applied Sciences, Beijing University of Technology, Beijing 100124, China \\
              \email{zhaohui@bjut.edu.cn}           
           \and
           N.H. Jing \at
             Department of Mathematics, North Carolina State University, Raleigh, NC 27695, USA \\
             Department of Mathematics, Shanghai University, Shanghai 200444, China
             \and
           S.M. Fei \at
             School of Mathematical Sciences, Capital Normal University, Beijing 100048, China
           }

\date{Received: date / Accepted: date}

\maketitle

\begin{abstract}
We generalize the notion of unextendible maximally entangled basis from bipartite systems to multipartite quantum systems. It is proved
that there do not exist unextendible maximally entangled bases in three-qubit systems. Moreover,
two types of unextendible maximally entangled bases are constructed in tripartite quantum systems and proved
to be not mutually unbiased.
\keywords{Unextendible maximally entangled bases \and Mutually unbiased bases \and Multipartite quantum systems}
\end{abstract}

\section{Introduction}
\label{intro}
Quantum entanglement has played an important role in various quantum information processings such as quantum teleportation [1], quantum cryptography [2], quantum dense coding [3], and parallel computing [4]. As one of the intrinsic features of quantum computation
and information, quantum
entanglement is closely related
to some of the fundamental problems in quantum mechanics such as reality and non-locality [5]. An important issue concerns with the
notion of unextendible product basis (UPB), which is a set of incomplete orthonormal product basis whose complementary space has no product states [6]. Moreover,UPBs play a rather diverse and important role in quantum information theory. Using the notion of UPBs a family of indecomposable linear maps had been built [7]. Bound entangled states were constructed based on the idea of UPBs [8]. UPBs can also demonstrate Bell inequalities without a quantum violation [9].

Corresponding to UPB is 
the unextendible maximally entangled basis (UMEB). It is known that there is no UMEB in the two-qubit system [10]. The UMEB in bipartite systems was studied and some explicit constructions of UMEB were given in Refs.[11,12,13]. In addition, the UMEB problem was generalized to states with given Schmidt numbers [14,15]. Though the bipartite case is well understood, the question of UMEB for multipartite systems is still an open problem.

Another related interesting notion is that of mutually unbiased base (MUB), which also plays an important role in quantum information. The maximum number of MUB in $\Bbb C^d$ is known to be no more than $d+1$ for any given $d$. It has been confirmed that there are indeed $d+1$ MUBs when $d$ is a prime power [16]. However, for general $d$, the maximum number of MUB is still open.

In this paper, we study UMEB in three-qubit system and MUB in tripartite systems. The paper is organized as follows: In Sect.2, we first generalize UMEB in bipartite systems to multipartite systems, then prove that there does not exist UMEB in three-qubit systems. In Sect.3, we construct different UMEBs in tripartite systems and show they are not mutually unbiased. Conclusions and discussions are given in Sect.4.

\section{UMEB in $\Bbb C^2\otimes \Bbb C^2\otimes \Bbb C^2$}
\label{sec:2}

We first recall the definition of UMEB in the bipartite system. Let $\Bbb C^{d_\alpha}$ be the $d_\alpha$-dimensional complex vector space. A set of states $\{|\phi_i\rangle\in \Bbb C^{d_1}\otimes \Bbb C^{d_2},\, i=1,2,\dots,n,\, n<d_1 d_2\}$ is called an $n$-member UMEB if and only if

$(\romannumeral1)$ $|\phi_i\rangle, i=1,2,\dots,n$, are maximally entangled;

$(\romannumeral2)\,\, \langle\phi_i|\phi_j\rangle=\delta_{ij}$;

$(\romannumeral3)$ If $\langle\phi_i|\varphi\rangle=0$ for all $i=1,2,\dots,n$, then $|\varphi\rangle$ cannot be maximally entangled.

Therefore the key component of the above UMEB is the concept of maximally entanglement of the bipartite system. To generalize the notion of the UMEB
to multipartite systems, we first recall the definition of maximally entanglement in multipartite situation. As the way of characterizing multipartite entanglement is not unique, different definition captures
different features of this quantum phenomenon. In this paper, we employ the definition of maximally multipartite entangled states introduced by Facchi et al [17].
For the existence of such maximally entangled states for qubit systems, see [18].

Consider a bipartition $(A,\bar{A})$ of system S, where $A\subset S$, $\bar{A}=S\setminus A$, $S=\{1,2,\dots,n\}$ and $1\le n_A\le n_{\bar{A}}$, with $n_A=|A|$, the cardinality of party A. At the level of Hilbert spaces, we get $H=H_A\otimes H_{\bar{A}}$.

\textbf{Definition 1:}
State $\rho\in H_A\otimes H_{\bar{A}}$, $N_A=dim(H_A)\le dim(H_{\bar{A}})$ is maximally entangled if and only if the reduced state is maximally mixed under all possible bipartite partitions $(A,\bar{A})$:
\begin{equation}
\rho_A=Tr_{\bar{A}} (\rho)= \dfrac{I}{n_A},
\end{equation}
where $I$ is corresponding identity matrix.

We now present the definition of UMEB in multipartite systems.

\textbf{Definition 2:}
A set of states $\{|\phi_i\rangle\in \Bbb C^{d_1}\otimes\ \Bbb C^{d_2}\otimes\dots\otimes \Bbb C^{d_k},\, i=1,2,\dots,n,\, n<d_1 d_2\dots d_k\}$ is called an $n$-member UMEB if and only if

$(\romannumeral1)$ $|\phi_i\rangle, i=1,2,\dots,n$, are maximally entangled;

$(\romannumeral2)\, \langle\phi_i|\phi_j\rangle=\delta_{ij}$;

$(\romannumeral3)$ If $\langle\phi_i|\varphi\rangle=0$ for all $i=1,2,\dots,n$, then $|\varphi\rangle$ cannot be maximally entangled.

Next we focus on quantum states in $\Bbb C^2\otimes \Bbb C^2\otimes \Bbb C^2$. In Ref. [10] the authors proved that UMEB does not exist by completing the basis
in $\Bbb C^2\otimes \Bbb C^2$, i.e. constructing $4$ basis vectors. However, UMEB in $\Bbb C^2\otimes \Bbb C^2$ can not have $4$ vectors. By the similar method we study UMEB in three-qubit system.

\textbf{Theorem:}
UMEB does not exist in $\Bbb C^2\otimes \Bbb C^2\otimes \Bbb C^2$.

\textbf{Proof~~}
For three qubits, the maximally entangled states are local unitary equivalent to the GHZ state [17].
Without loss of generality, a basis vector $|\phi\rangle$ of UMEB can be represented by a linear operator $U\otimes V$ acting on $\Bbb C^2\otimes \Bbb C^2$ such that
\begin{equation}
|\phi\rangle=(I\otimes U\otimes V)|\phi_0\rangle,
\end{equation}
where $U, V$ are unitary matrices over $\Bbb C^2$, and $|\phi_0\rangle=\dfrac{1}{\sqrt2}(|000\rangle+|111\rangle)$.

We can construct eight vectors which are maximally entangled and orthogonal to each other.
\begin{eqnarray}
\nonumber&&|\phi_1\rangle=(I\otimes I\otimes I)|\phi_0\rangle,\\
\nonumber&&|\phi_{\alpha+1}\rangle=(I\otimes I\otimes \sigma_\alpha)|\phi_0\rangle,~~ \alpha=1,2,3\\
\nonumber&&|\phi_{\beta+4}\rangle=(I\otimes \sigma_\beta\otimes I)|\phi_0\rangle,~~ \beta=1,2\\
&&|\phi_{\gamma+6}\rangle=(I\otimes \sigma_1\otimes \sigma_\gamma)|\phi_0\rangle,~~ \gamma=1,2
\end{eqnarray}
by the Pauli spin matrices
\begin{equation}
\sigma_1=
    \left(
      \begin{array}{ccccc}
        0 & 1 \\[2mm]
        1 & 0 \\[2mm]
      \end{array}
    \right)
,\  \sigma_2=
    \left(
      \begin{array}{ccccc}
        0 & -i \\[2mm]
        i & 0 \\[2mm]
      \end{array}
    \right)
, \ \ \sigma_3=
    \left(
      \begin{array}{ccccc}
        1 & 0 \\[2mm]
        0 & -1 \\[2mm]
      \end{array}
    \right).
\end{equation}

Any three-qubit pure state can be generally written in the form [19]:
\begin{equation}
|\varphi\rangle=\lambda_0|000\rangle+\lambda_1e^{i\theta}|100\rangle+\lambda_2|101\rangle+\lambda_3|110\rangle+\lambda_4|111\rangle,
\end{equation}
where $\lambda_i\ge0$, $\sum\limits_i \lambda_i^2=1$ and $\theta\in[0,\pi]$.
Next we will prove that if $\langle\phi_i|\varphi\rangle=0$ for all $i=1,2,\dots,8$, then $|\varphi\rangle$ cannot be maximally entangled.

Suppose $|\varphi\rangle$ is maximally entangled, consider $|\varphi\rangle$ as a three-qubit system associated to qubits A, B and C.
Under the bipartition $A|BC$, the reduced state of $|\varphi\rangle$ is of the form:
\begin{equation}
\begin{aligned}
  \rho_A = \lambda_0^2|0\rangle\langle0|+\lambda_0\lambda_1 e^{-i\theta}|0\rangle\langle1|
  +\lambda_0\lambda_1 e^{i\theta}|1\rangle\langle0|+(\lambda_1^2+\lambda_2^2+\lambda_3^2+\lambda_4^2)|1\rangle\langle1|,
\end{aligned}
\end{equation}
where $I$ is the identical operator in $H_A$. Setting  $\rho_A = \dfrac{I}{2}$, we get
\begin{equation}\label{b1}
\lambda_0^2=\dfrac{1}{2},~~\lambda_2^2+\lambda_3^2+\lambda_4^2=\dfrac{1}{2},~~\lambda_1=0.
\end{equation}

Let
\begin{equation}
U=
    \left(
      \begin{array}{ccccc}
        u_{11} & u_{12} \\[2mm]
        u_{21} & u_{22} \\[2mm]
      \end{array}
    \right),~~~
V=
    \left(
      \begin{array}{ccccc}
        v_{11} & v_{12} \\[2mm]
        v_{21} & v_{22} \\[2mm]
      \end{array}
    \right).
\end{equation}

Using Eq.(\ref{b1}) and $\langle\phi_i|\varphi\rangle=0$ for all $i=1,2,\dots,n$, we have
\begin{eqnarray}
\nonumber&&\langle\phi_1|\varphi\rangle=\dfrac{1}{\sqrt2}(\lambda_0 u_{11}v_{11}+\lambda_2 u_{21}v_{22}+\lambda_3 u_{22}v_{21}+\lambda_4 u_{22}v_{22})=0,\\
\nonumber&&\langle\phi_2|\varphi\rangle=\dfrac{1}{\sqrt2}(\lambda_0 u_{11}v_{21}+\lambda_2 u_{21}v_{12}+\lambda_3 u_{22}v_{11}+\lambda_4 u_{22}v_{12})=0,\\
\nonumber&&\langle\phi_3|\varphi\rangle=\dfrac{i}{\sqrt2}(\lambda_0 u_{11}v_{21}-\lambda_2 u_{21}v_{12}-\lambda_3 u_{22}v_{11}-\lambda_4 u_{22}v_{12})=0,\\
\nonumber&&\langle\phi_4|\varphi\rangle=\dfrac{1}{\sqrt2}(\lambda_0 u_{11}v_{11}-\lambda_2 u_{21}v_{22}-\lambda_3 u_{22}v_{21}-\lambda_4 u_{22}v_{22})=0,\\
\nonumber&&\langle\phi_5|\varphi\rangle=\dfrac{1}{\sqrt2}(\lambda_0 u_{21}v_{11}+\lambda_2 u_{11}v_{22}+\lambda_3 u_{12}v_{21}+\lambda_4 u_{12}v_{22})=0,\\
\nonumber&&\langle\phi_6|\varphi\rangle=\dfrac{i}{\sqrt2}(\lambda_0 u_{21}v_{11}-\lambda_2 u_{11}v_{22}-\lambda_3 u_{12}v_{21}-\lambda_4 u_{12}v_{22})=0,\\
\nonumber&&\langle\phi_7|\varphi\rangle=\dfrac{1}{\sqrt2}(\lambda_0 u_{21}v_{21}+\lambda_2 u_{11}v_{12}+\lambda_3 u_{12}v_{11}+\lambda_4 u_{12}v_{12})=0,\\
&&\langle\phi_8|\varphi\rangle=\dfrac{i}{\sqrt2}(\lambda_0 u_{21}v_{21}-\lambda_2 u_{11}v_{12}-\lambda_3 u_{12}v_{11}-\lambda_4 u_{12}v_{12})=0.
\end{eqnarray}
Hence
\begin{equation}
u_{11}v_{11}=u_{11}v_{21}=u_{21}v_{11}=u_{21}v_{21}=0.
\end{equation}
This result contradicts to the unitarity of $U$ and $V$, then $|\varphi\rangle$ is not maximally entangled. Therefore we can complete the basis $|\phi_1\rangle,\dots,|\phi_8\rangle$. But UMEB can not have $8$ vectors in $\Bbb C^2\otimes \Bbb C^2\otimes \Bbb C^2$ because of Definition $2$. Hence UMEB does not exist in $\Bbb C^2\otimes \Bbb C^2\otimes \Bbb C^2$.

\section{MUB in $\Bbb C^2\otimes \Bbb C^3\otimes \Bbb C^3$}
\label{sec:3}

MUB and UMEB have significant applications in quantum information processing. The relation between them is beginning to take notice. In this section we construct two UMEBs in $\Bbb C^2\otimes \Bbb C^3\otimes \Bbb C^3$ which are not mutually unbiased.

\textbf{Definition 3[11]:}
Two orthonormal bases $B_1=\{|b_i\rangle\}_{i=1}^d$ and $B_2=\{|c_j\rangle\}_{j=1}^d$ of $\Bbb C^d$ are said to be mutually unbiased if and only if
\begin{equation}
|\langle b_i|c_j\rangle|=\dfrac{1}{\sqrt d},~~~ \forall i,j=1,\dots,d.
\end{equation}


According to Ref.[11], we have two types of UMEBs in $\Bbb C^2\otimes \Bbb C^3$. One is
\begin{eqnarray}
\nonumber&&|\phi_0\rangle=\dfrac{1}{\sqrt2}(|00\rangle+|11\rangle),\\
&&|\phi_i\rangle=(\sigma_i\otimes I_3)|\phi_0\rangle,~ i=1,2,3.
\end{eqnarray}
Another is
\begin{equation}
|\psi_j\rangle=\dfrac{1}{\sqrt2}(\sigma_j\otimes I_3)(|0\rangle|x\rangle+|1\rangle|y\rangle),~~j=0,1,2,3,
\end{equation}
where $|x\rangle=\dfrac{1}{\sqrt3}(|0\rangle+\dfrac{1+\sqrt3i}{2}|1\rangle+|2\rangle)$,
$|y\rangle=\dfrac{1}{\sqrt3}(\dfrac{-\sqrt3+i}{2}|0\rangle+i|1\rangle-i|2\rangle)$ and
$\sigma_0=I$.

We now adopt the method in Ref.[15]. {Suppose $\{|\psi_i\rangle\}$ is an unextendible entangled bases with Schmidt number $k$ of $\Bbb C^{d_1}\otimes \Bbb C^{d_2}$,
where $d_1\leq d_2$, $1\leq k \leq d_1$, $|\psi_i\rangle=\sum\limits_{l=0}^{k}\lambda_l^{(i)}|\psi^{(i)}_l\rangle$,
and $|\psi^{(i)}_l\rangle=|a_l^{(i)}\rangle|b_l^{(i)}\rangle$ with $\{|a_l^{(i)}\rangle: l=1, \ldots, d_1\}$ an orthonormal basis of subsystem $\Bbb C^{d_1}$, and $\{|b^{(i)}_l\rangle: l=1, \ldots, d_2\}$ an orthonormal basis of subsystem $\Bbb C^{d_2}$.} If all Schmidt coefficients are equal to $\frac{1}{\sqrt{k}}$ and $k=d_1$, an unextendible entangled bases with Schmidt number $k$ reduces to UMEB.
Let
\begin{equation}
|\psi_{i,j}\rangle=\sum\limits_{l=0}^{d_1-1}\lambda_l^{(i)}|\psi^{(i)}_l\rangle|j\oplus l\rangle,
\end{equation}
where $\{|j\rangle\}$ is the standard computational basis of $\Bbb C^{d_3}$, $j=0,1,\dots,d_3-1$, $j\oplus l$ means $j+l$ mod $d_3$.
Then $\{|\psi_{i,j}\rangle\}$ is an UMEB of $\Bbb C^{d_1}\otimes \Bbb C^{d_2}\otimes \Bbb C^{d_3}$, $d_1\leq d_2 \leq d_3$ [15]. We can obtain two UMEBs in $\Bbb C^2\otimes \Bbb C^3\otimes \Bbb C^3$.

The first one is
\begin{eqnarray}
\nonumber&&|\phi_{0,0}\rangle=\dfrac{1}{\sqrt2}(|000\rangle+|111\rangle),\\
\nonumber&&|\phi_{0,1}\rangle=\dfrac{1}{\sqrt2}(|001\rangle+|112\rangle),\\
\nonumber&&|\phi_{0,2}\rangle=\dfrac{1}{\sqrt2}(|002\rangle+|110\rangle),\\
\nonumber&&|\phi_{i,0}\rangle=\dfrac{1}{\sqrt2}(\sigma_i\otimes I_3\otimes I_3)(|000\rangle+|111\rangle),\\
\nonumber&&|\phi_{i,0}\rangle=\dfrac{1}{\sqrt2}(\sigma_i\otimes I_3\otimes I_3)(|001\rangle+|112\rangle),\\
&&|\phi_{i,0}\rangle=\dfrac{1}{\sqrt2}(\sigma_i\otimes I_3\otimes I_3)(|002\rangle+|110\rangle),
\end{eqnarray}
where $i=1,2,3$.

The second one is
\begin{eqnarray}
\nonumber&&|\psi_{j,0}\rangle=\dfrac{1}{\sqrt2}(\sigma_j\otimes I_3\otimes I_3)(|0\rangle|x\rangle|0\rangle+|1\rangle|y\rangle|1\rangle),\\
\nonumber&&|\psi_{j,1}\rangle=\dfrac{1}{\sqrt2}(\sigma_j\otimes I_3\otimes I_3)(|0\rangle|x\rangle|1\rangle+|1\rangle|y\rangle|2\rangle),\\
&&|\psi_{j,2}\rangle=\dfrac{1}{\sqrt2}(\sigma_j\otimes I_3\otimes I_3)(|0\rangle|x\rangle|2\rangle+|1\rangle|y\rangle|0\rangle),
\end{eqnarray}
where $j=0,1,2,3$.

Due to $\langle\phi_{0,0}|\psi_{0,0}\rangle=\dfrac{1}{\sqrt6}$, the sets $\{|\phi_{i,l}\rangle\}$ and $\{|\psi_{j,l}\rangle\}$ are not MUBs in $\Bbb C^2\otimes \Bbb C^3\otimes \Bbb C^3$.

\section{Conclusion}
\label{sec:4}
We first have generalized the notion of UMEB from bipartite systems to multipartite quantum systems. Based on this, we prove that there does not exist an UMEB in $\Bbb C^2\otimes \Bbb C^2\otimes \Bbb C^2$. Moreover, we have constructed two types of UMEBs in $\Bbb C^2\otimes \Bbb C^3\otimes \Bbb C^3$ which are not mutually unbiased. Our results
may shed light on further investigation on UMEBs and MUBs for multipartite quantum states.

\begin{acknowledgements}
This work supported by the National Natural Science Foundation of China grant Nos. 11675113, 11281137, 11271138, 11101017 and 11531004 and Simons Foundation grant No.198129.
\end{acknowledgements}



\end{document}